\documentclass{article}
\usepackage{graphicx}
\usepackage{booktabs}
\usepackage{caption}
\graphicspath{{./IMG/}}
\usepackage{multicol}
\usepackage{multirow}
\usepackage{blindtext}
\usepackage{algorithm}
\usepackage{tipa}
\usepackage[utf8]{inputenc}
\usepackage[backend=biber, style=ieee,]{biblatex}
\usepackage{layout}
\usepackage{amsmath}
\usepackage{wrapfig}
\usepackage[a4paper, total={6in, 8in}]{geometry}
\usepackage{subcaption}
\usepackage{enumitem}
\usepackage[affil-sl, auth-lg]{authblk}
\usepackage{float}

\title{{\textbf{Non-locally averaged pruned reassigned spectrograms: a tool for glottal pulse visualization and analysis}\\}}

\author[a]{\textbf{Gabriel J. Griswold}\thanks{{Corresponding author}}}
\author[b]{\textbf{Mark A. Griswold}}
\affil[a]{ggriswol@usc.edu, Departments of Music Composition and Linguistics, University of Southern California, Los Angeles, CA, 90007}
\affil[b]{mark.griswold@case.edu, Departments of Radiology and Biomedical Engineering, Case Western Reserve University, Cleveland, OH, 44106}

\date{23 September 2025}

\begin{document}
\maketitle

\begin{abstract}
Reassigned spectrograms have shown advantages in precise formant measuring and inter-speaker differentiation. However, reassigned spectrograms suffer from their inability to visualize larger amounts of data($>$40ms) in an easily comprehensible and reproducible manner. Utilizing the techniques and tools developed by Fulop and Fitz, a variation of the reassigned spectrogram is proposed. Non-locally Averaged Pruned Reassigned Spectrograms (NAPReS) provide a simplified view into the characteristics of a speaker’s glottal pulsation patterns throughout the centroid of a vowel through the stacking, summing, and pruning of large numbers of glottal pulses. In this exploratory study, NAPReS has been shown to display a large amount of data in an easily comprehensible and quantifiable manner, while also making the observation of low-amplitude cyclical structures more accessible. NAPReS also allows for alternative formant fitting methods such as Gaussian mixture modeling (GMM). In this study, NAPReS with GMM was compared against conventional LPC fitting of formant values and was shown to be more reproducible than conventional LPC fitting in high-noise situations.
\par
\textbf{Keywords:} \par
Reassigned spectrograms, Non-local averaging, Glottal pulse analysis, Voice differentiation, Forensics
\par
\textbf{Highlights:}
\begin{enumerate}
    \item Introduces NAPReS, a modified approach to glottal pulse analysis utilizing reassigned spectrograms
    \item Show application of Gaussian Mixture Modeling (GMM) to NAPReS
    \item Show advantages of NAPReS in high-noise environments
\end{enumerate}
\end{abstract}

\section{Introduction}
\subsection{Overview}
 
Time-corrected instantaneous frequency spectrograms (TCIFs), more commonly referred to as reassigned spectrograms, first proposed by Kodera et al. in 1976~\cite{Kodera1976ANM}, serve as a tool to visualize very short non-stationary multi-component FM signals in a higher resolution and fidelity than is possible with conventional spectrograms. Multiple methods for the computational rendering of such representations have been proposed independently by Kodera, Nelson~\cite{Nelson}, and Auger/Flandrin~\cite{AugerFlandrin}. 
\par
Fulop and Fitz pioneered the technique for linguistic use by utilizing Nelson’s methods for cross-spectral processing of speech in order to create precise representations of signal components while unifying the theoretical framework proposd in prior studies ~\cite{Fulop2004AlgorithmsFC,FulopFitz2006}. 
Reassigned spectrograms are able to precisely represent the frequency and amplitude of a component signal across a short timescale. This can allow for a more nuanced understanding of the previously indistinguishable characteristics of glottal pulses.
\par
The implementation of reassigned spectrograms for use in forensic speaker identification was first proposed by Fulop and Disner in 2007~\cite{fulopanddisnertoolforvoiceid}. Through further publication by Fulop and Disner~\cite{FulopDisnerforensicID2}, the usefulness of reassigned spectrograms in forensic settings has become even clearer. Reassigned spectrograms have been shown to produce different representations of glottal pulsation patterns across multiple speakers. As Fulop and Disner noted, the irregularity in the mean structure of pulses throughout an utterance represents the largest individuating factor \cite{fulopanddisnertoolforvoiceid}. Because of this, while reassigned spectrograms cannot definitively identify the speaker, they can help to distinguish individuals when it comes to visual comparison in a digital quasi-voice lineup \cite{fulopanddisnertoolforvoiceid,FulopDisnerforensicID2}. Utilizing reassigned spectrograms in this setting has, however, only been hypothesized in the literature and has rarely been utilized in real-world forensic settings. 
\par 
A significant part of this lack of utilization comes from the fact that reassigned spectrograms struggle to visualize large amounts of temporal data in a readable fashion. Most reassigned spectrograms only visualize an approximately 40 ms partial utterance and are themselves difficult to contextualize. It is difficult to understand what aspects of the signal repeat, and what aspects of the signal occur only in a few pulses which may, or may not have been sampled. An example of left-to-right paired pruned reassigned spectrograms can be seen in Figure 2 of S. Fulop and S. Disner \cite{fulopanddisnertoolforvoiceid}.
\par
Here, we aim to achieve two distinct goals. First, we provide an automated method for non-local averaging to provide concise, human-readable reassigned spectrograms of a subject's glottal pulses. Second, we use the improved averaging to provide an increased signal-to-noise ratio (SNR), particularly in high-noise environments. 

\par
Different forms of spectral analysis with simplification and/or denoising have previously been applied in multiple areas outside of linguistics with varying levels of automation. For example, Baudes et al \cite{NLmeans} proposed using non-local averaging to denoise images by identifying common regions of an image and averaging them to improve the visual appearance of an image. Similarly Jabloun et al \cite{EEG} have applied non-local averaging to electroencephalogram (EEG) spectra, demonstrating a clear improvement in visualization of spectra peaks in the EEG waveforms. Other denoising approaches without non-local averaging include using wavelet denoising to improve the apparent SNR of either seismic data \cite{Chen2014} or radar data \cite{Stankovic2004}.  All of these methods are outside of linguistics but have shown potential benefits of applying optimized averaging methods to the visualization of structured spectral data. 

\subsection{Reassignment Process}
The reassignment procedure entails a multi-stage computational framework wherein energy is relocated from geometric time-frequency centers to positions that more precisely represent signal characteristics (Auger and Flandrin, 1995). This process begins with standard Short-Time Fourier Transform computation (STFT). In a conventional spectrogram calculated from an STFT, we place energy at the center point $(t,\omega)$ of each time-frequency box on our grid. This is equivalent to saying "all the energy belongs exactly at this central point," however this doesn’t take into account the full context of the signal. The reassignment method estimates where the energy actually belongs. It's like looking at a blurry photograph and sharpening it by moving each bit of light to where it truly originated from. Mathematically this is accomplished by calculating two auxiliary transforms using window function derivatives to determine the corrective displacement vectors. Where $X(\omega,t)$
 represents the standard STFT of the signal using the analysis window, and $X_t(\omega,t)$ denotes the STFT computed using the time-derivative of the analysis window, this involves computing two corrections, one for time given by: 

\begin{equation}
\hat{t}(t,\omega) = t - \text{Re}\left\{\frac{X_t(t,\omega)}{X(t,\omega)}\right\}
\end{equation}

and another for frequency:

\begin{equation}
\hat{\omega}(t,\omega) = \omega + \text{Im}\left\{\frac{X_\omega(t,\omega)}{X(t,\omega)}\right\}
\end{equation}

which systematically relocates spectral energy according to local phase behavior (Fulop and Fitz, 2006). These two calculations take into account the "instantaneous frequency" - the exact frequency at each moment (rather than assuming it's at the center of our frequency bin) and the "group delay" - the precise time when each frequency component actually occurred (rather than assuming it's at the center of our time window). By moving energy from the geometric centers to these calculated positions, one can often obtain a sharper picture of the true signal content, especially for signals that change quickly over time or contain multiple components.

 \par However, the algorithm must address potential instabilities in regions where signal energy approaches zero or where the phase information becomes unreliable. To avoid these areas, the reassigned spectrogram implementation used here employs a pruning mechanism based on Nelson's (2001) indicator functions to filter out unreliable reassignment estimates. First, an amplitude threshold eliminates low-energy components that are susceptible to phase instability, retaining only time-frequency points where the signal magnitude exceeds a user-defined decibel threshold relative to the maximum (which was taken as -100dB in this work.) Second, frequency boundary constraints restrict analysis to a user-specified spectral region of interest, such as between 100Hz and 10000Hz for speech and music applications. Third, temporal containment criteria ensure that reassigned time coordinates remain within the boundaries of the analyzed signal, preventing edge artifacts. The algorithm also applies phase derivative stability tests that evaluate the mixed partial derivatives of both instantaneous frequency and group delay. Points are retained only when either the absolute value of the instantaneous frequency derivative is less than 0.25 (indicating stable frequency trajectories) or when the absolute deviation of the group delay derivative from unity is less than 0.25 (indicating reliable time localization). These differential criteria effectively detect regions where the signal's phase behaves in a physically meaningful manner, corresponding to genuine acoustic components rather than interference patterns or numerical instabilities. The combined application of these criteria produces a sparse yet highly informative representation that highlights coherent signal components while suppressing artifacts and noise.

\subsection{Theory}
 Unlike these previous methods, we aim to improve the performance of reassigned spectrograms to both summarize longer patterns of glottal pulses and to increase the sensitivity of the summarized results. This method, Non-locally Averaged Pruned Reassigned Spectrograms (NAPReS), can automatically condense long segments of glottal pulses into a single-pulse reassigned spectrogram. At the same time, the sensitivity of this representation is increased compared to previous methods by averaging multiple aligned pulses. This allows for simple representations of complex acoustical environments, which reveal precise formant structures that were previously difficult to observe across the timescale represented. It also provides opportunities to implement alternative spectral analysis methods.
\par
NAPReS, to explain simply, analyzes groups of consecutive glottal pulses from an utterance and automatically slices them into individual pulses. A manual selection of glottal pulses is required, however once this has been done, NAPReS automatically determines the location of the onset of each pulse. Once the number and duration of each glottal pulse has been independently recorded, their similarity to one another is assessed, and points of similarity are defined between pulses. Following this, the individual slices are then aligned and summed in a process derived from non-local means~\cite{NLmeans}. The data are then pruned according to Nelson’s method~\cite{Nelson} but derived from the code produced by Fulop and Fitz~\cite{FulopFitz2006}.  The resulting reassigned spectrogram does not depict any particular pulse, as in previous studies, rather the resulting NAPReS depicts the mathematically derived mean structure of the glottal pulses in the selected utterance. 

\section{Mathematical basis for NAPReS}
\subsection{Mathematical overview}
\par
The process for creating a NAPReS spectrogram begins with defining a template pulse that will be used to align all subsequent pulses in the selected utterance. The template pulse's length is derived from the fundamental frequency of the utterance and represents approximately one complete opening and closing of the subject's vocal folds, though more than one pulse may be used if desired. Once this template pulse has been selected, NAPReS searches through the rest of the input signal for patterns which match the template pulse. This is possible due to the underlying assumption that all consecutive glottal pulses share a similar spectrum. In order to achieve template matching, the inner product of the template pulse and each shifted position of signal is calculated. Within this vector of inner products, each peak represents the onset of a pulse which corresponds to the onset of the template pulse. Once the set of matching pulses is defined, they are averaged together with the template pulse to form the composite spectrogram which represents the values, not of one particular pulse, but rather the average pulse in the selected utterance. From here the composite spectrogram is passed through a reassignment and pruning process, exactly as described by Nelson~\cite{Nelson} and Fulop/Fitz~\cite{Fulop2004AlgorithmsFC,FulopFitz2006}.

\begin{figure}[H]
    \centering
\includegraphics[scale = .5]{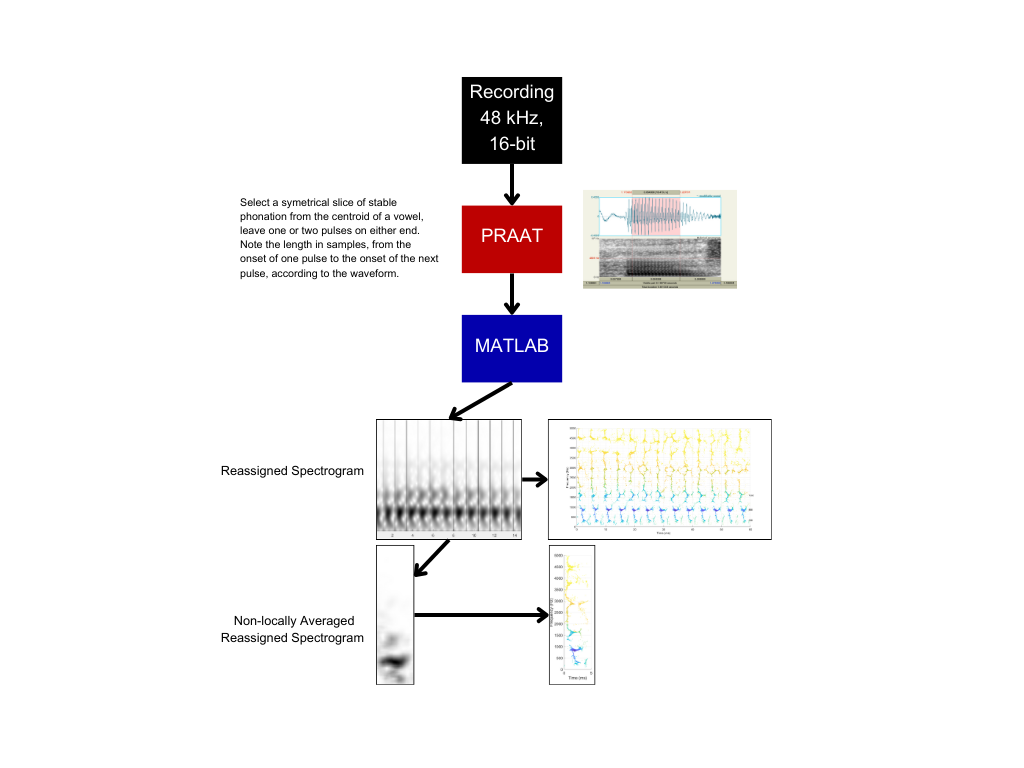}
\centering
\caption{A schematic representing the process of creating a NAPReS.}
\label{fig:schematic}
\end{figure}

\subsection{Process}
\par
The input waveform is converted to a short-time Fourier transform (STFT) spectrogram as described in Nelson~\cite{Nelson}. Here a STFT of length N is defined as a discrete function $f[n]$. For all of the work shown here, a length of 2048 samples was used and a Hanning window was applied prior to STFT, where the Hanning window is defined as:

\begin{equation}
\mathbf{w}(n) = 0.5 \left(1 - \cos\left(\frac{2\pi n}{w-1}\right)\right), \quad n = 0, 1, \ldots, w-1
\end{equation}

We start with the assumption that the first segment is the template and compare shifted spectra to this position with a fixed width of $w$ samples. In this case, a spectral region can be defined as $f[m]$ for $m \in [m, m+w]$. The inner product $p[i]$ between the template and the shifted spectrum is defined as:

\begin{equation}
p[i]=\sum_{n=1}^{N}\left| f[n] \right| \left|f[n+i]\right|
\end{equation}

for $i=1...N$. Note that the magnitude of the spectrogram is used to perform this matching as it is more stable than the complex-valued calculation.
\par
Let $P_J$ be the set of all peaks found in the function $p[i]$:

\begin{equation}
p[i - 1] < p[i] \geq p[i + 1]
\end{equation}

We define $J$ to be the total number of pulses found in the signal.

\par The width of the template pulse is defined as $w$ samples. In this case, the set of matching spectral regions can be defined as $f[m]$ for $m \in [m, m+w]$ and where $m \in P_J$. Thus we can average the spectral regions together to form: 

\begin{equation}
f_\text{avg}[n]=\sum_{i\in P_J}^{}\frac{f[i]}{J}
\end{equation}

Following this non-local averaging process, the spectrum is then passed through the framework described by Fulop and Fitz~\cite{Fulop2004AlgorithmsFC} which results in a NAPReS image. Please note that this algorithm assumes that the input signal is properly cropped to the desired analysis region.

\section{Alternative Analysis Methods}
\par
Linguistic analysis of spectrograms often centers on the frequencies of the formants. Linear predictive coding (LPC) has been a well-established gold standard because of its ease of computation and validated performance. However, it is well known in many spectrographic applications that LPC has poor performance in low signal-to-noise ratio (SNR) situations, and can fail to find all of the relevant peaks in complicated spectra \cite{Pitch-SynchronousLPC}. Because the preprocessing provided by NAPReS results in a simplified spectral representation with robust sensitivity to higher frequency formants, alternative spectral analysis methods may be beneficial. In this study, we have explored the use of Gaussian Mixture Models (GMM)\cite{FormantPaperGaussian}. GMM is well known in spectrographic analysis for its ability to robustly represent multiple spectral peaks even at low SNR values. In GMM, the spectrum is represented as a weighted sum of multiple Gaussian components:

\begin{equation}
S(f) = \sum_{i=1}^{M} A_i \exp\left(-\frac{(f - \mu_i)^2}{2\sigma_i^2}\right)
\end{equation}

for $M$ Gaussian components. Like conventional reassigned spectrograms, NAPReS results in a series of point results that are displayed as a complete spectrogram. In order to fit these individual points to a GMM, we first make a histogram of the NAPReS spectrogram along the frequency dimension. This 1D spectrum is then fit to a GMM to extract the peak locations. There are multiple ways to solve for these types of models. In this case, we used the Matlab built-in function "fit()" with either five or six components for this purpose depending on the number of formants visible in the spectrum. 

\begin{figure}[H]
    \centering
\includegraphics[scale = .6]{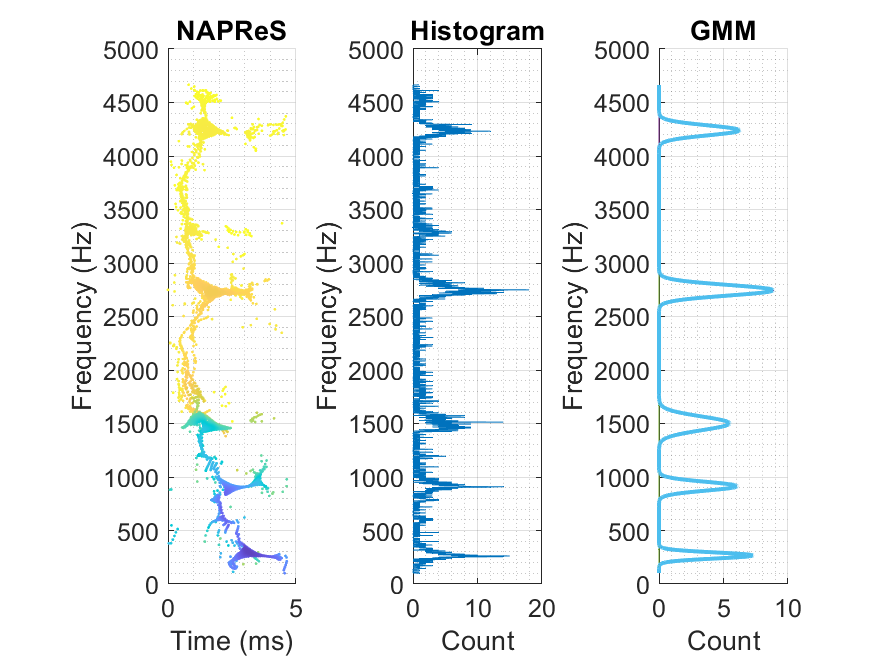}
\centering
\caption{An example of the use of a GMM to fit a NAPReS spectrum. The points from the NAPReS spectrum (left) are projected into a 1D spectrum by calculating the histogram (middle). This historgram is then fit to the GMM, resulting in a composite fitted spectrum, as shown on the right, which now clearly depicts the formant locations. }
\label{fig:GMM_schematic}
\end{figure}

In the following figures, the peaks found in each NAPReS spectrogram using GMM are displayed on the right hand side of each figure.

\section{Applying NAPReS}
\subsection{Utterance selection}
Each subject was instructed to speak the following utterances:

\setenumerate[0]{nosep}
\begin{enumerate}
    \item "Password" three times in their normal, unmodified, modal voice.
    \item "Uh, Um, Un" [\textipa{\textturnv, \textturnv  m, \textturnv  n].}
\end{enumerate}

\subsection{Data sources}
\vspace{1pt}
\par
Six Individuals participated in this study after providing written informed consent. Recordings were made using an iPhone 15 Pro at a 48 kHz sampling rate. All recordings were free of audible background noise and recorded in an acoustically controlled environment. The audio files were transferred to secure cloud storage and subsequently processed on a personal computer. Each utterance was segmented into glottal pulse groups using PRAAT, and both conventional reassigned spectrograms and NAPReS representations were computed in MATLAB. The resulting spectrograms were exported in standard image formats for analysis and visualization. 
\par In addition, the authors contributed their own recordings under identical conditions. For transparency, they self-identify here as Subject 6 (S6) and Subject 8 (S8).

\section{Data}
\subsection{Intraspeaker variation}

\begin{figure}[H]
    \centering
\includegraphics[scale = .75]{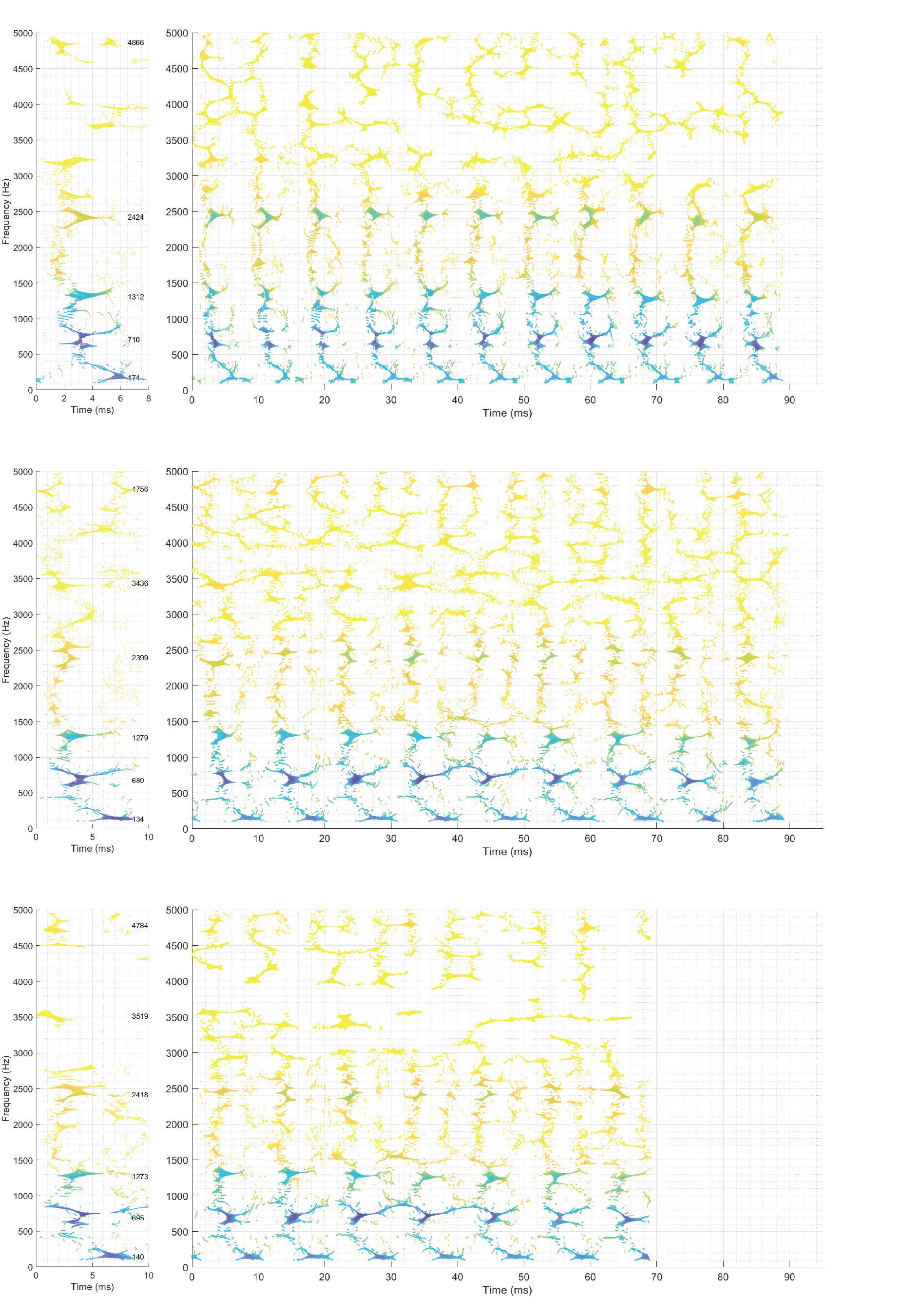}
\centering
\caption{NAPReS images for three isolated [\ae] utterances by speaker 7 with  corresponding traditional reassigned spectrograms}
\label{fig:S7_triple}
\end{figure}

\begin{figure}[H]
    \centering
\includegraphics[scale = .8]{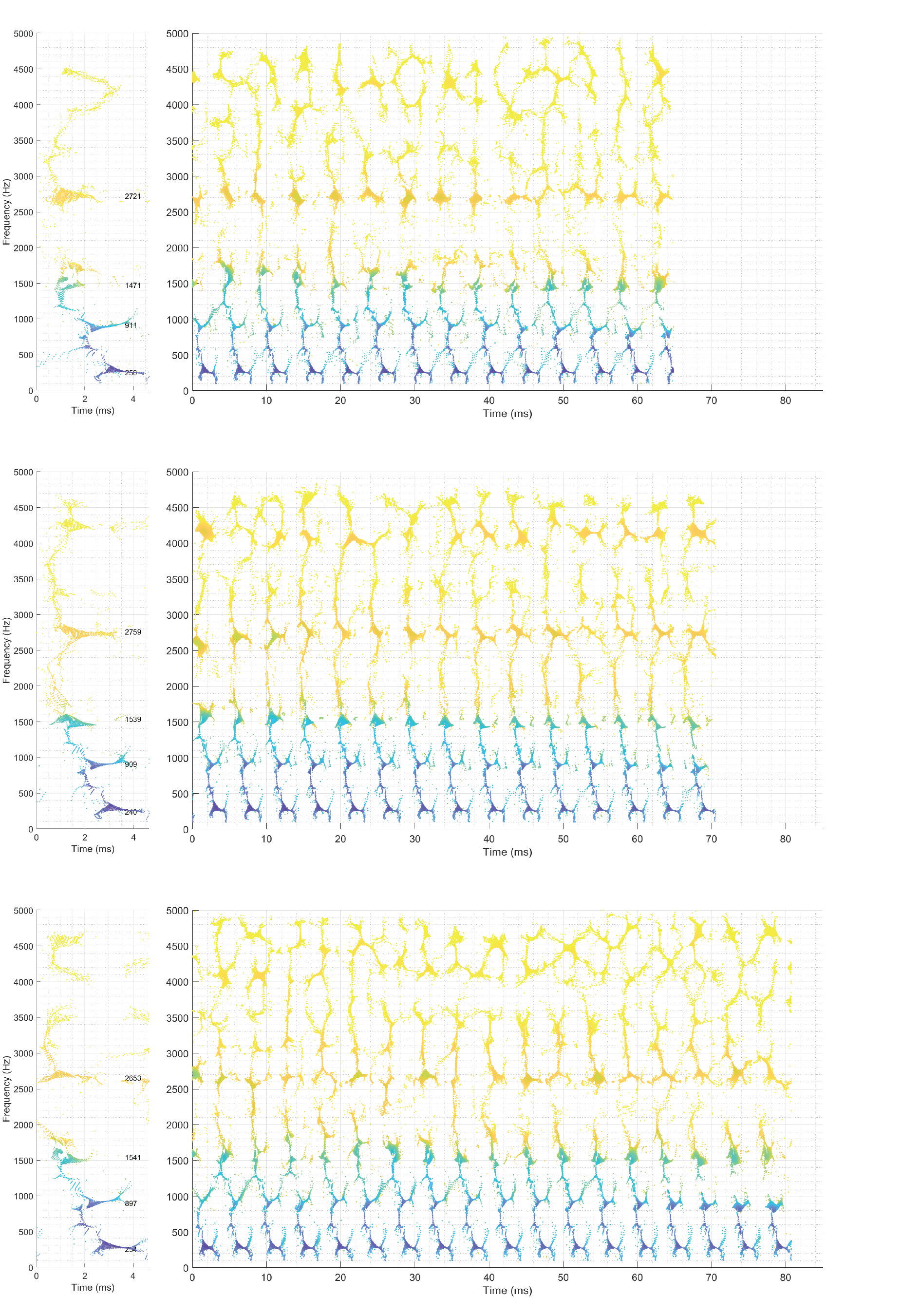}
\centering
\caption{NAPReS images for three isolated [\ae] utterances by speaker 1 with  corresponding traditional reassigned spectrograms}
\label{fig:S1_triple}
\end{figure}

Figures \ref{fig:S7_triple} and \ref{fig:S1_triple} juxtapose reassigned spectrograms (right) with their NAPReS counterparts (left). As shown when comparing these two methods, NAPReS condenses a large number of ambiguous pulses into a singular representation which displays the data contained in that utterance. Speakers 7 and 1 display intraspeaker congruence with all three NAPReS visualizations appearing visually similar in structure, amplitude, and possessing computer-determined formant values that remain similar across utterances.
\par 
Speaker 7 provided three utterances which, without any direction to do so, decreased in $F_0$ over time. The $F_0$ variance middle and bottom utterances in Figure \ref{fig:S7_triple} is 24 Hz. Comparing these most disparate utterances in figure \ref{fig:S7_triple} provides insights into the extent to which structures remain similar with a wide speaker-specific $F_0$ variance.
\par
In both of these figures, one can observe how NAPReS clarifies the underlying acoustical structure of these pulses while removing non-cyclical, noisy signals. For instance, in the middle NAPReS of figure \ref{fig:S1_triple}, one can observe a clearly defined $F_3$ at 2759Hz. In the traditional reassigned spectrogram to the right the $F_3$ structure and frequency shows much more irregularity, most likely due to its lower relative amplitude. NAPReS demonstrates a clear formant structure while also, as in the middle NAPReS of figure \ref{fig:S7_triple}, removing noisy signals. In this and other NAPReS images,  particularly in frequencies higher than $F_3$, but also in the region between $F_2$ and $F_3$, the non-cyclical noise is greatly reduced in amplitude because the signal in these bands is incoherent from pulse to pulse compared to the repeating signals one expects to occur in formant-producing frequency bands. Together, both qualities lead to NAPReS's ease of visualization when analyzing a sample of equal or greater duration than a traditional reassigned spectrogram. 

\subsection{NAPReS at varying $J$-values}
\begin{figure}[H]
    \centering
\includegraphics[scale = .6]{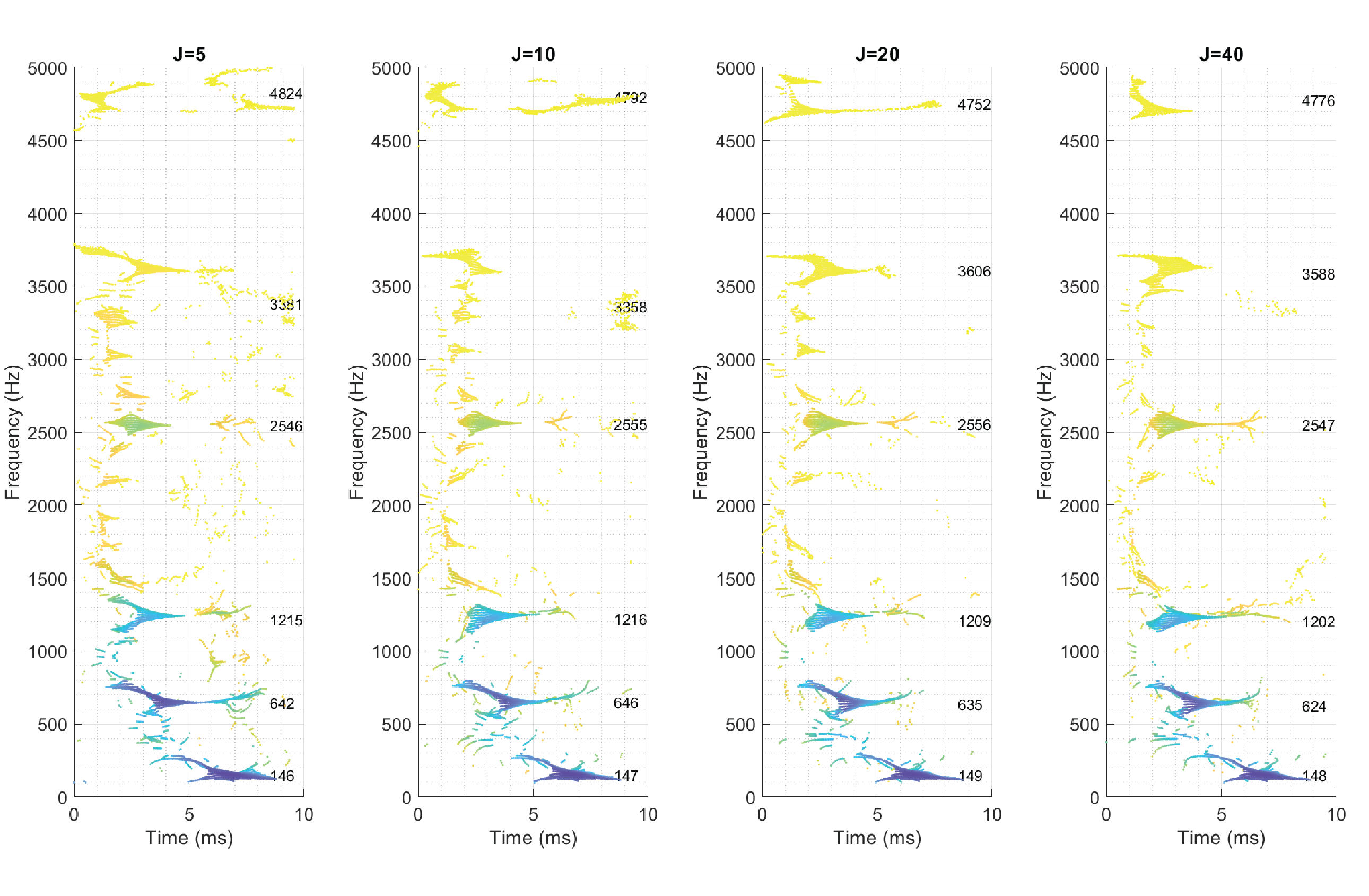}
\centering
\caption{One utterance of [\textturnv] by S7 analyzed at varying $J$-values ranging from $J$ = 5 to $J$ = 40}
\label{fig:S7_jvalues}
\end{figure}

\begin{figure}[H]
    \centering
\includegraphics[scale = .40]{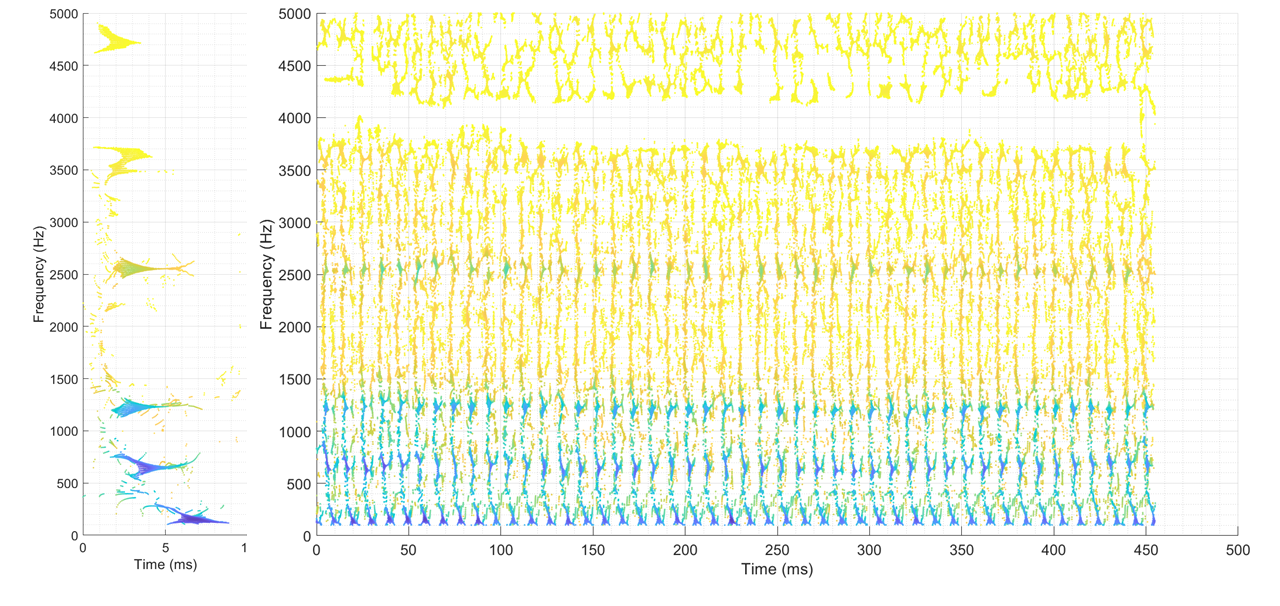}
\centering
\caption{The $J=40$ utterance from figure \ref{fig:S7_jvalues} as well as the entirety of the 47 pulse pruned reassigned spectrogram for the same utterance.}
\label{fig:S7_j_full}
\end{figure}

\par
NAPReS is capable of analyzing any utterance at a variety of $J$-values. As shown in figures \ref{fig:S7_jvalues} and \ref{fig:S7_j_full}, adding more glottal pulses to the NAPReS visualization results in slightly different structures. $J$-values exceeding 15-20 should not be expected when analyzing normal speech patterns; however $J$-values are naturally different from speaker to speaker as a result of that speaker's $F_0$. For low speakers, very few pulses may be attainable, and the opposite is true for speakers with a higher $F_0$. In analyzing these structures, our dataset implies that past $J$ = 20 there is little significant change in either the amplitude, frequency, or visual appearance of any of the formants present in S7’s speech sample. These nearly half-second long NAPReS images with significantly higher $J$-values are possible only because of the speaker’s slower speech cadence. It is clear in analyzing these images that there is a point at which the $J$-value becomes relatively stable, however, more analysis is required in order to understand the concreteness of this boundary, as well as whether NAPReS images below this hypothetical $J$-boundary should be rejected from analysis. 

\subsection{Interspeaker variation}
\begin{figure}[H]
    \centering
\includegraphics[scale = 1]{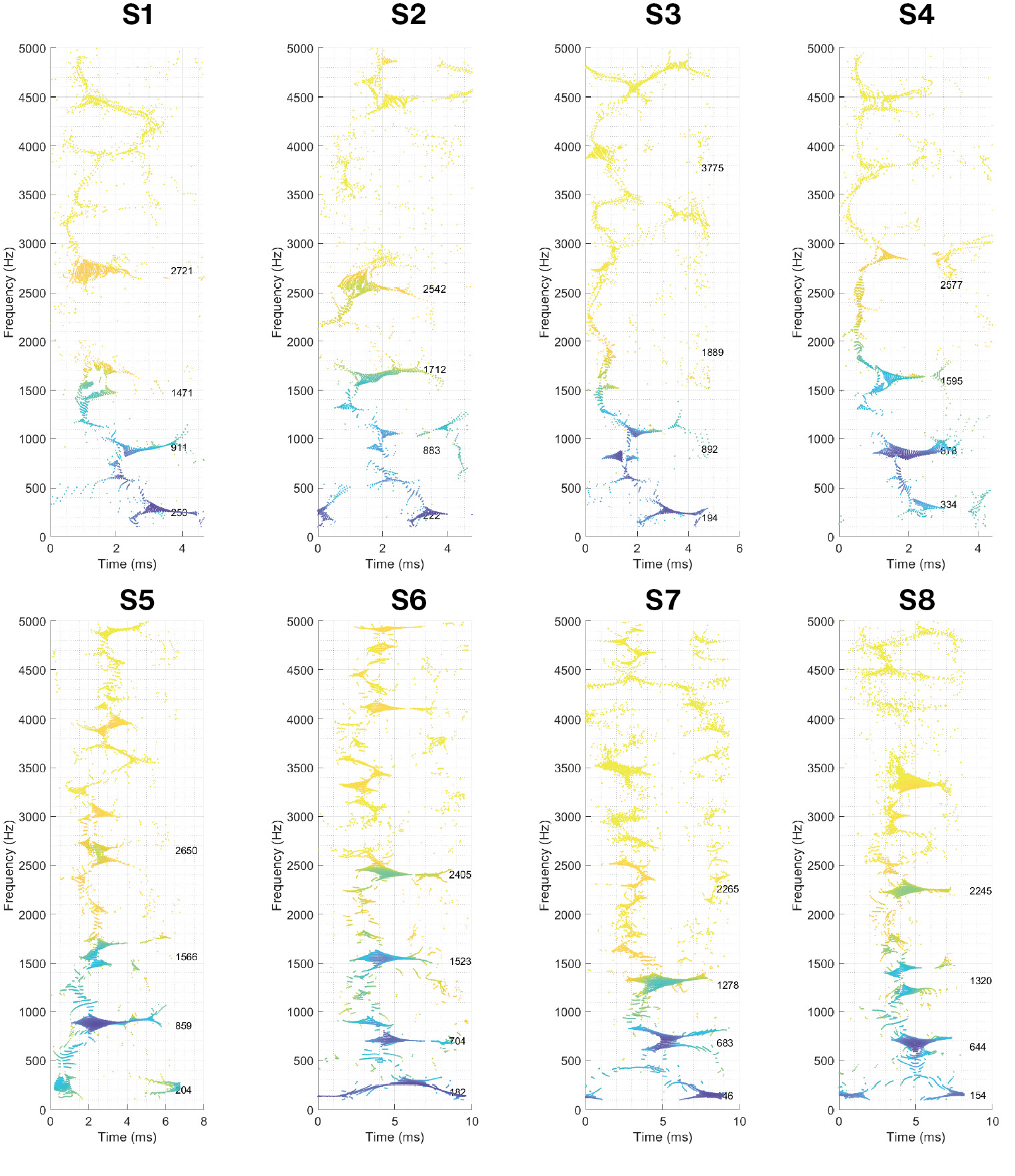}
\centering
\caption{Eight [\ae] utterances by all subjects. The lowest $J$-value in this figure is $J=9$ by S6 and the highest is $J=14$ by S1.}
\label{fig:group}
\end{figure}

\par
In Figure \ref{fig:group} we can see NAPReS images for each speaker. In every speaker except speaker 3, NAPReS was able to compute a precise formant up to $F_3$. S3 speaks with a very soft and airy voice which may have added to NAPReS's inability to compute or identify such low-amplitude repetition. To counter this, S3's voice may need to be analyzed with a lower clipping value, used to remove excess noise, however, for this publication, all voices were analyzed at a respectively uniform clip value. 
\par
Speakers S1 and S4 speak in a similar manner, dialect, and cadence, and with an $F_0$ variance of only 3Hz, yet S4's NAPReS does not visually represent, in any way, any of the three NAPReS images presented in figures \ref{fig:S1_triple} or \ref{fig:group}.
\par

\subsection{Sensitivity to Noise}
The non-local averaging intrinsic to NAPReS provides the potential to analyze audio waveforms with lower SNR as well as the analysis of higher formant frequencies as compared to more conventional methods. To this end, we performed a Monte Carlo analysis of NAPReS with GMM fitting as compared to conventional LPC for the identification of formant frequencies. We started with the long utterance from S7 used above and added varying amounts of white noise. The root-mean-squared (RMS) intensity of this waveform was 0.2 (a.u.) and thus we added white noise to give final SNR values relative to this baseline of 100, 5, 2 and 1. Twenty replicas of the signal plus white noise were created and stored as a WAV files. Conventional LPC formant locations were calculated using PRAAT's Gaussian-like analysis window with a length of 25ms with the window beginning at the first glottal onset. 67 separate estimates were generated across the width of the entire utterance, and all of these 67 resulting formant values were averaged together across all 20 replicas. Similarly, NAPReS results were calculated using the entire utterance followed by GMM fitting as described above, using 6 spectral components.
\par
The results of this analysis are summarized in Table 1. At high SNR, both methods provide similar results for $F_1$-$F_4$, though NAPReS is able to achieve stable results also for $F_5$. At the lower SNRs tested, both methods show decreased performance. However, NAPReS continues to provide stable mean values all the way down to SNR=1, although with clearly increasing uncertainties. On the other hand, LPC in PRAAT shows both an increase in uncertainty as well as a significant bias towards higher formant frequencies. At the lower SNRs, the errors are so significant that the ranges of the low SNR LPC results no longer overlap with the true value as determined by the spectra of the original recording. Please note that Table 1 also indicates the number of fit failures for each method. In these tests, LPC showed a higher failure rate at lower SNR as compared to GMM. The statistical measures displayed are only derived from successful fits, and thus there is a bias towards lower standard deviation. All of the remaining results could be improved by removing outliers from the results at these lower SNRs, but we have chosen to display all of the results to give a full indication of the performance of both methods.     

These results indicate that NAPReS may be useful in situations where the source audio quality is low or for the analysis of higher formants, which typically have much lower SNR than the lower frequency formants. 

\begin{table}[ht!]
\centering
\caption{Monte Carlo noise-sensitivity results for NAPReS and LPC (F1--F5). Values reported are Mean, Std, Min, Max, and Failures at four SNR levels.}
\begin{tabular}{lcccc | lcccc}
\toprule
 & \multicolumn{4}{c}{NAPReS} & & \multicolumn{4}{c}{LPC} \\
\cmidrule{2-5} \cmidrule{7-10}
 & SNR100 & SNR5 & SNR2 & SNR1 & & SNR100 & SNR5 & SNR2 & SNR1 \\
\midrule
\multicolumn{9}{l}{\textbf{F1}} \\
Mean     & 653.3 & 653.4 & 651.4 & 652.3 & & 654.8 & 673.5 & 875.6 & 964.2 \\
Std      & 0.5   & 1.2   & 3.6   & 6.9   & & 0.1   & 1.1   & 13.7  & 10.8  \\
Min      & 652.8 & 650.5 & 644.1 & 641.0 & & 654.6 & 669.9 & 853.7 & 950.0 \\
Max      & 654.6 & 655.7 & 655.8 & 665.5 & & 655.1 & 675.2 & 900.7 & 987.1 \\
Failures & 0     & 0     & 0     & 0     & & 0     & 0     & 0     & 0     \\
\midrule
\multicolumn{9}{l}{\textbf{F2}} \\
Mean     & 1228.0 & 1228.8 & 1231.3 & 1230.7 & & 1224.3 & 1224.8 & 1317.0 & 1965.6 \\
Std      & 0.4    & 1.5    & 3.4    & 9.3    & & 0.1    & 1.1    & 56.7   & 75.5   \\
Min      & 1227.5 & 1225.5 & 1224.3 & 1209.2 & & 1224.2 & 1222.7 & 1221.4 & 1793.0 \\
Max      & 1229.0 & 1231.9 & 1241.2 & 1245.4 & & 1224.5 & 1227.1 & 1465.0 & 2067.5 \\
Failures & 0      & 0      & 0      & 0      & & 0      & 0      & 0      & 0      \\
\midrule
\multicolumn{9}{l}{\textbf{F3}} \\
Mean     & 2552.5 & 2552.7 & 2554.5 & 2525.5 & & 2535.8 & 2559.6 & 2780.5 & 3081.6 \\
Std      & 0.8    & 3.9    & 22.9   & 353.0  & & 0.3    & 3.0    & 48.3   & 53.0   \\
Min      & 2550.8 & 2543.7 & 2492.2 & 1919.3 & & 2535.4 & 2554.0 & 2702.7 & 2958.1 \\
Max      & 2553.7 & 2558.9 & 2594.0 & 3362.7 & & 2536.5 & 2564.7 & 2911.0 & 3191.8 \\
Failures & 0      & 0      & 0      & 0      & & 0      & 0      & 0      & 0      \\
\midrule
\multicolumn{9}{l}{\textbf{F4}} \\
Mean     & 3606.8 & 3590.3 & 3529.1 & 3613.5 & & 3477.7 & 3641.4 & 3979.2 & 4211.3 \\
Std      & 6.4    & 38.3   & 387.5  & 213.2  & & 1.3    & 12.3   & 25.0   & 43.5   \\
Min      & 3588.8 & 3496.2 & 2736.8 & 3100.5 & & 3475.4 & 3612.0 & 3934.2 & 4117.0 \\
Max      & 3614.9 & 3659.6 & 4767.8 & 4087.7 & & 3480.4 & 3657.9 & 4010.3 & 4309.1 \\
Failures & 0      & 0      & 1      & 0      & & 0      & 0      & 11      & 4      \\
\midrule
\multicolumn{9}{l}{\textbf{F5}} \\
Mean     & 4727.5 & 4774.0 & 4609.7 & 4556.1 & & -- & -- & -- & -- \\
Std      & 25.7   & 63.3   & 332.9  & 226.7  & & --     & --     & --     & --     \\
Min      & 4699.8 & 4682.2 & 3697.8 & 4016.9 & & --     & --     & --     & --     \\
Max      & 4806.2 & 4924.5 & 5166.7 & 4823.6 & & --     & --     & --     & --     \\
Failures & 0      & 3      & 6      & 3      & & 20     & 20     & 20     & 20     \\
\bottomrule
\end{tabular}
\end{table}

\section{Discussion}
Here we have presented a novel method, NAPReS, for processing and visualizing glottal pulses for linguistic analysis using non-local averaging of reassigned spectrograms. The method automatically identifies and aligns slices of a conventional audio recording based on the similarity of their glottal pulse patterns. The resulting visualization represents approximately 3-4 times more data (12-16 glottal pulses) than could previously be shown using a reassigned spectrogram. The results enable quick and easy assessment of formant height, relative amplitude, and relative structure. 
\par
The inherent averaging in NAPReS also provides improved sensitivity in high-noise environments. This allows for the application of different analysis approaches. Here, we compared the fitting of formant frequencies using both conventional LPC within PRAAT and GMM using NAPReS. The GMM results showed significantly better performance at low SNRs. While both methods could be tuned to different situations, the ability to use GMM in a straightforward way offers several advantages over LPC. In general, GMM provides greater flexibility by representing the spectrum as a weighted sum of Gaussian distributions. This allows GMM to better capture complex, multi-modal frequency distributions with overlapping peaks that LPC might smooth over. While we did not perform this analysis here, GMM can also model both the positions and shapes of spectral peaks simultaneously, adapting to varying peak widths and asymmetries. It should be noted that pitch-synchronous LPC methods with noise compensation such as \cite{Pitch-SynchronousLPC}, show similarly degraded precision in higher formants under low SNR conditions as our LPC results show here. The NAPReS results, while based on a small dataset, illustrate a more stable representation of higher-frequency resonances by averaging across glottal cycles. This suggests that NAPReS potentially addresses a persistent limitation of LPC highlighted in prior work \cite{Pitch-SynchronousLPC}.  
\par
NAPReS is not limited to just the simplistic GMM-based analyses performed here. Other non-GMM analysis methods are also possible using the condensed representations from NAPReS. As an example, data were shown here across multiple subjects that support NAPReS's potential use in voice differentiation. At the end of the day, this becomes an image recognition problem, and the ongoing revolution in AI/machine learning methods has shown significant gains in this kind of problem over the last few years. It's likely that these methods could directly benefit from the preprocessing simplification and reduced input matrix size provided by NAPReS.
\par
The current study has several limitations. First, in the current implementation, NAPReS requires a manual preprocessing step to extract a single phoneme and to identify the glottal pulse frequency. These can be performed using conventional methods, which are quick and efficient, but future implementations could also automate these steps. It is also clear that both the conventional LPC and GMM fitting used here are relatively simplistic and that alternative methods are possible \cite{Pitch-SynchronousLPC}\cite{matthews1961pitch}\cite{Parthasarathy}. However, even these other methods should benefit from the compressed, higher SNR representation from NAPReS. The NAPReS results shown here were also only from a limited set of subjects. Future quantitative studies using larger cross-sections of subjects need to be performed in order to validate the potential use cases described here.

\section{Conclusions}
\par
NAPReS is a tool for the visualization and analysis that provides a single, condensed view of a series of glottal pulses. NAPReS has the ability to improve sensitivity by removing non-cyclical features of the signal, while simultaneously reinforcing the core acoustical elements of the glottal pulses. 

\section*{Acknowledgements}
We would like to thank Dr. Sean Fulop of California State University Fresno for his discussions on reassigned spectrograms and for providing portions of code which have been adapted for use in this research. Many thanks as well to Dr. Sandra Ferrari Disner of the University of Southern California for her mentorship, teaching, and discussion relating to this paper. We also thank Dr. Khalil Iskarous of the University of Southern California for his guidance in preparing this manuscript.

\section*{Author Declarations}
The authors have no conflicts to disclose. 

\section*{Data availability statement}
The data that support the findings of this study are available from the corresponding author upon reasonable request.
\printbibliography
\end{document}